\documentclass[pdflatex,sn-nature]{sn-jnl}
\usepackage{graphicx}%
\usepackage{multirow}%
\usepackage{amsmath,amssymb,amsfonts}%
\usepackage{amsthm}%
\usepackage{mathrsfs}%
\usepackage[title]{appendix}%
\usepackage{xcolor}%
\usepackage{textcomp}%
\usepackage{manyfoot}%
\usepackage{booktabs}%
\usepackage{algorithm}%
\usepackage{algorithmicx}%
\usepackage{algpseudocode}%
\usepackage{listings}%
\usepackage{gensymb}
\usepackage{tabularx}
\usepackage{anyfontsize}
\begin{document}

\title[Article Title]{Light Storage in Light Cages: A Scalable Platform for Multiplexed Quantum Memories}

\author*[1]{\fnm{Esteban} \sur{Gómez-López}}\email{egomez@physik.hu-berlin.de}

\author[1]{\fnm{Dominik} \sur{Ritter}}

\author[2]{\fnm{Jisoo} \sur{Kim}}

\author[3]{\fnm{Harald} \sur{K\"ubler}}

\author[2,4]{\fnm{Markus A.} \sur{Schmidt}}

\author[1]{\fnm{Oliver} \sur{Benson}}

\affil*[1]{\orgdiv{Department of Physics}, \orgname{Humboldt-Universit\"at zu Berlin}, \orgaddress{\city{Berlin}, \postcode{12489}, \country{Germany}}}

\affil[2]{\orgdiv{Department of Fiber Photonics}, \orgname{Leibniz Institute of Photonic Technology}, \orgaddress{\city{Jena}, \postcode{07745}, \country{Germany}}}

\affil[3]{\orgdiv{5. Physikalisches Institut and Center for Integrated Quantum Science and Technology}, \orgname{University of Stuttgart}, \orgaddress{ \city{Stuttgart}, \postcode{70569}, \country{Germany}}}

\affil[4]{\orgname{Otto Schott Institute of Material Research}, \orgaddress{\city{Jena}, \postcode{07743}, \country{Germany}}}

\abstract{
Quantum memories are essential for photonic quantum technologies, enabling long-distance quantum communication and serving as delay units in quantum computing. Hot atomic vapors using electromagnetically induced transparency provide a simple platform with second-long photon storage capabilities. Light-guiding structures enhance performance, but current hollow-core fiber waveguides face significant limitations in filling time, physical size, fabrication versatility, and large-scale integration potential.

In this work, we demonstrate the storage of attenuated coherent light pulses in a cesium (Cs) quantum memory based on a 3D-nanoprinted hollow-core waveguide, known as a light cage (LC), with several hundred nanoseconds of storage times. Leveraging the versatile fabrication process, we successfully integrated multiple LC memories onto a single chip within a Cs vapor cell, achieving consistent performance across all devices. We conducted a detailed investigation into storage efficiency, analyzing memory lifetime and bandwidth. These results represent a significant advancement toward spatially multiplexed quantum memories and have the potential to elevate memory integration to unprecedented levels. We anticipate applications in parallel single-photon synchronization for quantum repeater nodes and photonic quantum computing platforms.

}


\maketitle

\section{Introduction}\label{intro}

The ability to store quantum information is crucial for various quantum technologies. In long-distance quantum communication, signal loss in quantum channels poses a fundamental limitation \cite{kimble_quantum_2008}. Quantum repeaters \cite{briegel_quantum_1998} mitigate this challenge through entanglement swapping, where quantum memories play a key role by enabling teleportation and synchronizing entanglement-swapping operations \cite{azuma_quantum_2023,van_loock_extending_2020}. Experimental studies demonstrate that incorporating quantum memories in repeater segments significantly enhances photon pair rates compared to memory-less implementations \cite{pu_experimental_2021,davidson_single-photon_2023}. Beyond communication, quantum memories are essential in photonic quantum computing, providing photon synchronization and controllable delays necessary for feed-forward operations in measurement-based quantum computing \cite{barrett_scalable_2010,xu_demonstration_2012,davidson_single-photon_2023}.

Quantum memories have been developed across various platforms, including solid-state systems \cite{lago-rivera_telecom-heralded_2021,bussieres_quantum_2014,morton_solid-state_2008,berezhnoi_quantum_2022,knowles_controlling_2017,tang_storage_2015}, ultra-cold atoms \cite{bao_efficient_2012,cho_highly_2016,ding_toward_2014,chen_deterministic_2006,vernaz-gris_highly-efficient_2018}, and hot atomic vapors using electromagnetically induced transparency (EIT) \cite{katz_light_2018,wolters_simple_2017,dou_broadband_2018,buser_single-photon_2022,wang_field-deployable_2022,cho_atomic_2010,bashkansky_quantum_2012,vurgaftman_suppressing_2013,kaczmarek_high-speed_2018}. Hot atomic vapors offer low technical complexity, making them attractive for scalable quantum memories. Long storage times of hundreds of milliseconds have been demonstrated in cesium vapor \cite{katz_light_2018}. 

Combining atomic vapor quantum memories with waveguiding structures enhances light-matter interaction, reducing the memory's footprint and the optical power required to operate it. Recent experimental implementations have demonstrated the storage of light via electromagnetically induced transparency (EIT) using atoms within the core of optical waveguides, predominantly employing hollow-core fibers (HCFs) \cite{sprague_broadband_2014,blatt_stationary_2016,peters_single-photon-level_2020,leong_long_2020,li_controlled_2020,rowland_high-bandwidth_2024}.  However, these approaches face significant challenges. Loading atomic vapors into the cores of HCFs is typically done through the fiber's end facets, resulting in filling times that can span months \cite{sprague_broadband_2014,rowland_high-bandwidth_2024}. Alternatively, laser-cooled atoms have been used to load the fibers \cite{blatt_stationary_2016,peters_single-photon-level_2020,leong_long_2020,li_controlled_2020}, but this introduces additional experimental complexity due to the need for atom trapping and cooling. Moreover, the fiber-pulling fabrication process constrains the shape and design of hollow-core fiber memories. HCFs are not easily integrated into optical chips, further limiting their practicality.

To address these limitations, we focus on the unique versatility of so-called hollow-core light cages (LCs) to realize quantum memories with a small footprint, great flexibility in design, and easy integration into photonic chips. LCs are anti-resonant hollow core waveguides with unique side-wise access to the core region \cite{jain_hollow_2019}. These on-silicon-chip structures, 3D-nanoprinted from commercially available photoresins using two-photon polymerization, have shown their worth in optofluidic analysis \cite{kim_optofluidic_2021}, spectroscopy \cite{kim_-chip_2022} and nanoparticle characterization \cite{kim_locally_2022-1}. Additionally, the LCs show a notable enhancement in the diffusion of molecular gases \cite{jain_hollow_2019}, as well as atomic gases \cite{davidson-marquis_coherent_2021}, compared to the traditional HCFs, reducing the time needed to reach reasonable optical depths, from months to days, while keeping the optical fields confined in the LC core. By tuning the geometrical parameters of the LCs, the anti-resonances can be tuned to guide light across specific spectral intervals in the visible spectrum and the near-infrared \cite{jain_hollow_2019,jang_fiber-integrated_2021,kim_-chip_2022}.
Coherent light-matter interaction has been shown by creating EIT inside LCs placed in a Cs vapor cell and achieving Rabi frequencies in the order of hundreds of MHz \cite{davidson-marquis_coherent_2021}. 

In this work, we demonstrate the use of LCs as on-chip quantum memories based on electromagnetically induced transparency (EIT), achieving storage times of several hundred nanoseconds. We integrated multiple LC memories onto a single chip within a cesium (Cs) vapor cell and observed consistent performance across all devices. This marks a significant step toward spatially multiplexed quantum memories and could scale up memory integration to a new level. 
We anticipate applications in parallel single-photon synchronization for quantum repeater nodes and photonic quantum computers. Our thorough investigation includes the storage efficiency of faint coherent light pulses in LCs and an analysis of the memory's lifetime and bandwidth, emphasizing their dependence on control power during writing and reading processes. Moreover, we examine the current limitations of LC-based quantum memories, especially concerning efficiency and decoherence times.

\section{Results}\label{results}

\subsection{Light cages in atomic vapor}

\begin{figure}[ht]
    \centering
    \includegraphics[width=1.0\textwidth]{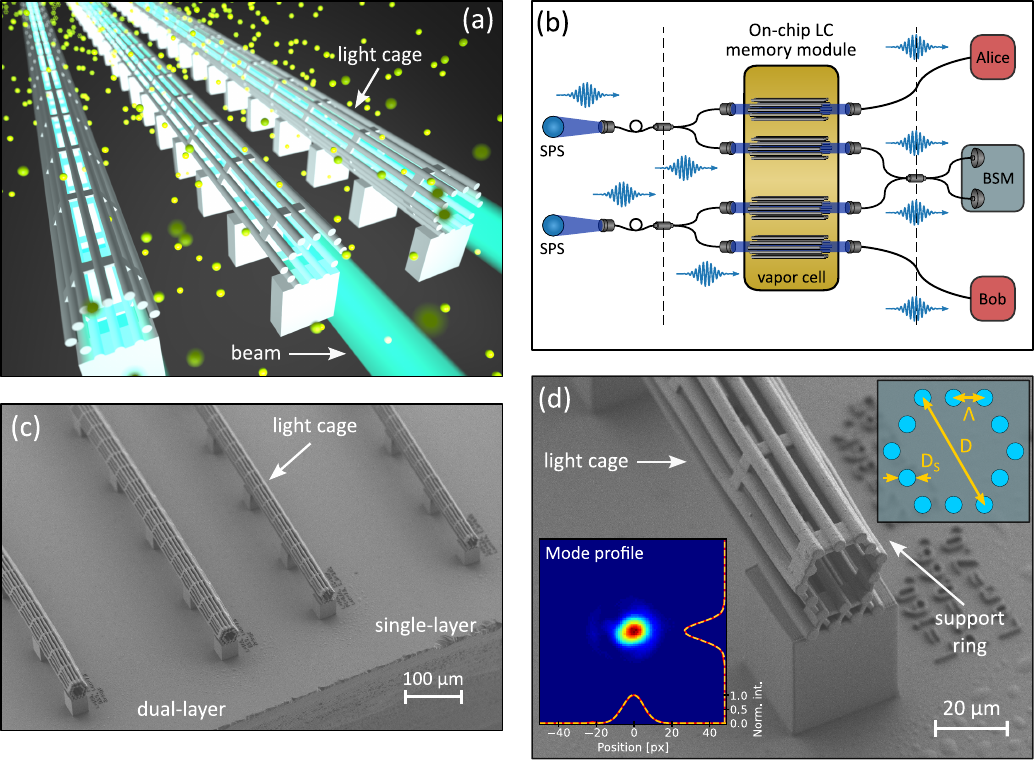}
    \caption{
    \textbf{Spatially multiplexed quantum memory using hollow-core light-cages (LC) on a chip.}
    \textbf{(a)} Illustration of several LC guiding light through their core immersed in a Cs atmosphere.
    \textbf{(b)} Scheme of the use of LCs as a spatially multiplexed on-chip quantum memory (see text). 
    \textbf{(c)} SEM image of four LCs printed on the same chip, with two different geometries, single-layer and dual-layer of strands. Printing several waveguides hundreds of micrometers apart in the same printing process is possible.    
    \textbf{(d)} Zoom-in of a 12-strand LC. \textbf{Top inset}: Cross section of the LC with the relevant design parameters, $D_s = 3.6~\mu m$, $\Lambda = 7~\mu m$, $D = 28~\mu m.$ \textbf{Bottom inset}: Collimated output beam. x-y beam profiles fitted to Gaussian functions (yellow dashed lines) showing the fundamental mode $\mathrm{TEM}_{00}$ is coupled in the LC.
    }
    \label{fig-concept}
\end{figure}

A LC is a special type of anti-resonant hollow-core waveguide \cite{zeisberger_analytic_2017} that guides light along its core while allowing for fast diffusion of Cs inside the core through its unique side-wise access \cite{davidson-marquis_coherent_2021}, as illustrated in Fig. \ref{fig-concept}-(a). The conceptual idea of a spatially multiplexed on-chip quantum memory is schematically shown in Fig. \ref{fig-concept}-(b). Flying qubits encoded in the polarization or time-bins of individual photons from single photon sources  (SPS) are separated by beam splitters (polarization beam splitters or switches) and fiber-coupled to several light cages (LCs) on a chip inside an atomic vapor cell. After storage and synchronization by choosing different storage times for each LC the photons are released, coupled into fibers, and partially processed, e.g., by a Bell state measurement (BSM). Such a functionality is needed, e.g., in quantum repeaters.     

The LCs were designed and 3D-nanoprinted on a silicon substrate following the two-photon polymerization technique described in ref. \cite{jain_hollow_2019} (see Methods). A scanning electron microscope (SEM) image in Fig. \ref{fig-concept}-(c) shows several fabricated LCs on a chip. A zoom-in of an individual LC and the measured mode profile of guided light is depicted in Fig. \ref{fig-concept}-(d). The LC chip is placed inside a Cs vapor cell, and light is coupled into individual LCs. To prevent degradation of the polymer LCs by the chemically reactive Cs gas, the structure is coated with 100 nm of alumina. We found remarkable stability and observed no indication of degradation even after 5 years in the Cs atmosphere. The coating also serves to fine-tune the waveguide resonances \cite{jang_fine-tuning_2020} to the Cs D1 line at $894~nm$. 

\begin{figure}[ht]
    \centering
    \includegraphics[width=1.0\textwidth]{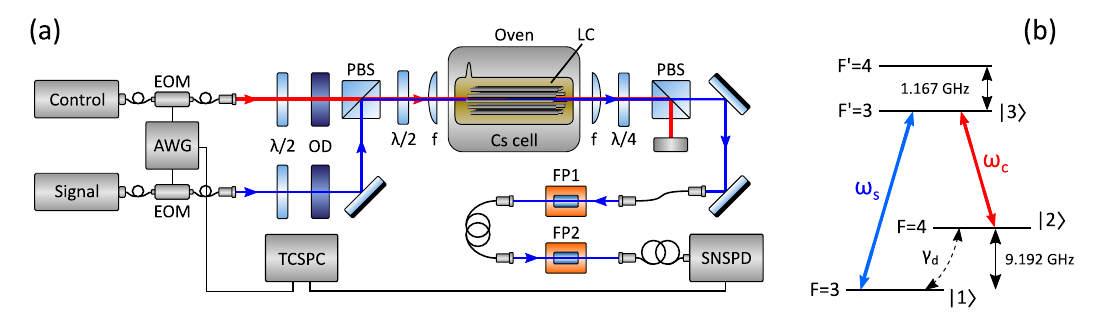}
    \caption{\textbf{Experimental setup.} \textbf{(a)} EIT is created using co-propagating signal and control fields, coupled into the LC with a pair of aspheric lenses (f). The control field is filtered using polarization and frequency filtering. Pulses are carved from CW lasers using a pair of EOMs. (OD: neutral density attenuators, $\Lambda/2$: half-wave plate, $\Lambda/4$: quarter-wave plate, PBS: polarizing beam splitter, FPx: tunable Fabry-Pérot cavities, EOM: electro-optic modulator, AWG: arbitrary waveform generator, SNSPD: superconducting nanowire single-photon detector, TCSPC: time-correlated single photon counting unit). \textbf{(b)} Energy level of the employed $\lambda$-system in the Cs D1 line. Signal ($\omega_s$) and control ($\omega_c$) fields are locked to the transitions F=3-F’=3 and F=4-F’=3. The decoherence rate between the ground states F=3 and F=4 is denoted as $\gamma_d$.}
    \label{fig-setup}
\end{figure}

The experimental setup of the quantum memory is shown in Fig. \ref{fig-setup}-(a). Signal and control laser beams are generated from independent diode lasers locked to the transitions $F=3-F'=3$ and $F=4-F'=3$ of the Cs D1 line, respectively (see energy level diagram in Fig. \ref{fig-setup}-(b)). The beams are coupled collinearly with orthogonal polarizations into the LCs on the chip inside the vapor cell using a pair of aspherical lenses. We found a coupling efficiency $\eta_{LC}=0.20(1)$. After transmission, the control beam is suppressed through polarization filtering, and the signal is detected by a superconducting nanowire single-photon detector (SNSPD). A key part of the memory, the vapor cell, is heated to $T=74~\degree C$ to increase its optical depth, and the transmission spectra are measured while scanning the signal laser frequency (see Methods for more details). 

\begin{figure}[ht]
    \centering
    \includegraphics[width=1.0\textwidth]{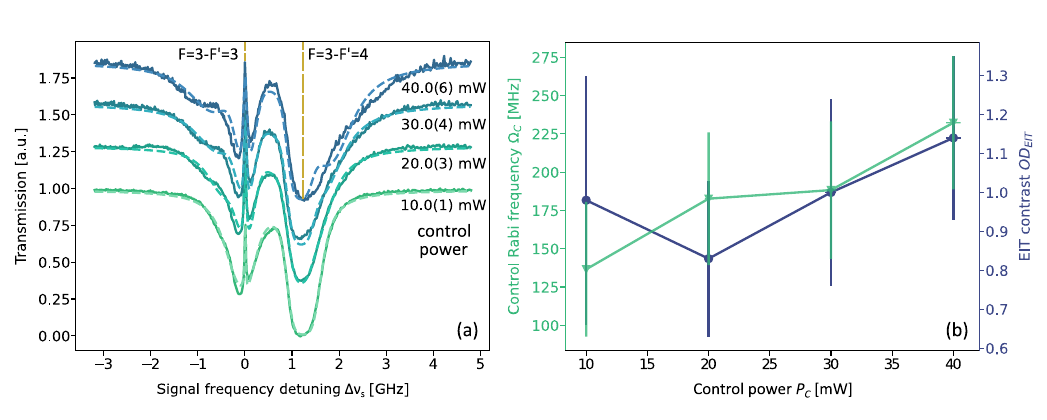}
    \caption{\textbf{EIT in the LC.} \textbf{(a)} Measured (solid line) EIT transmission spectra of light coupled into the LC at different control powers. The EIT peaks are visible on the two-photon resonance ($\Delta\nu_s=0$) with increasing widths and transmission. Theoretical fits are plotted as dashed lines. \textbf{(b)} The Rabi frequency and EIT contrast ($OD_\mathrm{EIT}$) as obtained from the fits at varying control powers (see details in the main text).}
    \label{fig-EITspectra}
\end{figure}

In Fig. \ref{fig-EITspectra}-(a), we can observe the EIT transmission peak at the two-photon resonance (signal detuning $\Delta\nu_s=0$) for different powers of the control beam. With increasing power, the transmission within the transparency window increases and approaches nearly unity at $P_c=40~mW$. The apparition of "shoulder-like" features at $\Delta\nu_s \sim -0.8~GHz$ and $1.6~GHz$ are caused by the propagation loss in the LC. These features confirm that the observed EIT stems from light propagation in a mode of the LC. The spectra are fitted by the model proposed in reference \cite{davidson-marquis_coherent_2021}, using a propagation-dependent linear susceptibility, $\chi(z)$, arising from the attenuation of the control Rabi frequency as $\Omega_c(z) = \Omega_c(0) e^{\alpha z}$, with $\alpha$ being the modal attenuation of the LC (see supplementary information \ref{SI-EIT-simulation}). Using this model, we can extract $\Omega_c$, which is plotted in Fig. \ref{fig-EITspectra}-(b), as a function of the control power. Here we observe an increasing $\Omega_c$ for rising control powers, showing a maximum value of $\Omega_c=232(43)~MHz$. The corresponding transparency window width $\Delta f_\mathrm{EIT}=133(24)~MHz$ is computed from $\Omega_c$ after considering the Doppler broadening and optical depth of the atomic medium. On the other hand, the EIT contrast ($OD_\mathrm{EIT}$), defined as the reduction in optical depth at the EIT resonance \cite{finkelstein_practical_2023}, remains almost constant with a value of $OD_\mathrm{EIT}\approx1$. This shows the LC is capable of handling high control laser powers producing broad transparency windows, a requirement for storage of broadband light pulses.

\subsection{Storage of light in a light cage}

To implement the EIT storage protocol, we cut pulses out of the continuous wave signal and control beams by a pair of amplitude electro-optic modulators (EOM). A first control pulse performs optical pumping. Then, a faint signal pulse with 2/3 the length of the control pulse is sent simultaneously with a second control pulse. This transfers the signal pulse into a spin wave in the atomic ensemble via EIT. After a set storage time $t_\mathrm{storage}$, the process is reversed, i.e., the stored light is released by sending a third control pulse. 

\begin{figure}[ht]
    \centering
    \includegraphics[width=1.0\textwidth]{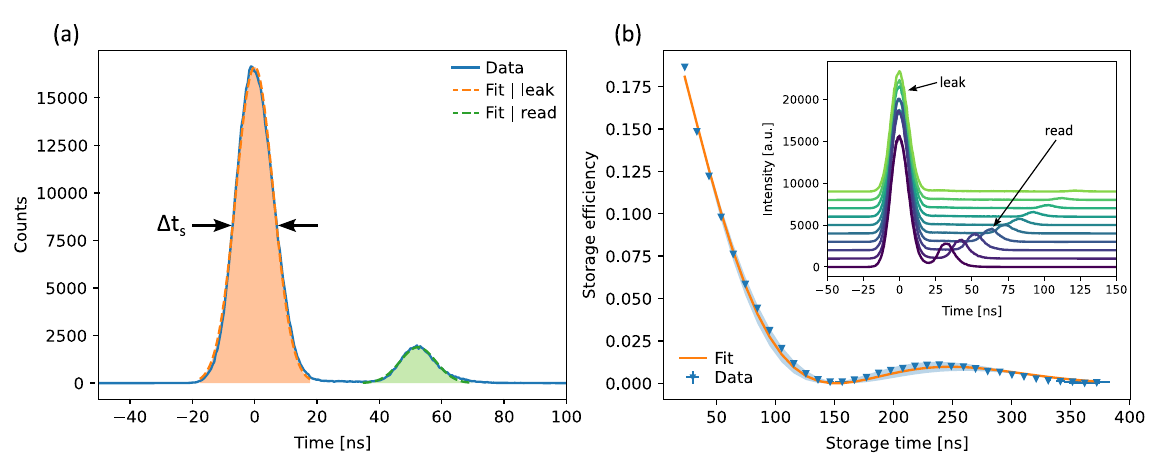}
    \caption{\textbf{Storage of light in the LC.} \textbf{(a)} The leak pulse (filled in orange), i.e., the part of the input pulse that escaped storage, sets the time zero. Arrows indicate the pulse width. The retrieved read pulse (filled in green) appears after $52~ns$. Experimental parameters are: $T = 74.0~\degree C$, input signal pulse width $\Delta t_{s} = 14.1(4)~ns$, coupling power $P_c=10.0(1)~mW$. The internal efficiency for $52~ns$ storage is $\eta_{int}=0.0982(1)$. \textbf{(b)} Memory lifetime measured for different storage times. A $1/e$-lifetime of $83(2)~ns$ is obtained using a heuristic fit that considers the precession of electronic spins produced by stray magnetic fields (see text). The shaded blue area indicates the 95\% confidence interval of the fit. \textbf{Inset}: Selected set of measurements analyzed to calculate the memory lifetime. An intensity offset is added for better visibility.}
    \label{fig-storage_analysis}
\end{figure}

Fig. \ref{fig-storage_analysis}-(a) shows a measurement using a peak control power of $P_c=10~mW$ and an attenuated signal pulse characterized by an average photon number of $|\alpha|^2=50(5)$ and a pulse width (FWHM) of $\Delta t_s = 14.1(4)~ns$. The control noise is subtracted by measuring the control field without the signal field. The first prominent peak from light that escaped storage ("leak") defines the time zero in the plot. We define the internal memory efficiency as the ratio of the integrated counts from the retrieved signal ("read") to the total amount of counts, $\eta_\mathrm{int}=N_{\mathrm{read}}/(N_\mathrm{read}+N_\mathrm{leak})$. A temporal filter of width $3\sigma_\mathrm{in}$ is applied to both pulses, where $\sigma_\mathrm{in}$ is the variance of the fitted Gaussian distribution to the leak pulse (shaded areas in Fig. \ref{fig-storage_analysis}-(a)). For a storage time of $52~ns$, the internal memory efficiency is $\eta_\mathrm{int}=0.098(1)$. This shows that the LC can store light as a spin-wave excitation in the atomic medium even in a regime where $OD_\mathrm{EIT}$ is relatively low.

We then studied the dependency of the memory efficiency $\eta_\mathrm{int}$ on the set storage time. The results are plotted in Fig. \ref{fig-storage_analysis}-(b). Surprisingly, the efficiency does not show the expected mono-exponential decay but a damped oscillatory behavior. We attributed this to a spin precession due to residual magnetic fields while the light is stored as a spin-wave. We fit the data using an exponentially damped sinusoidal oscillation, where the oscillation frequency is determined by a magnetic field. The fit suggests a magnetic field that agrees well with the field expected from the earth and heating wires (see supplementary information \ref{SI-Bfield-efficiency}). The observed spin precession enables a novel way of magnetic polarization control of the stored light pulses, an application that will be studied further in future works.

\begin{figure}[ht]
    \centering
    \includegraphics[width=1.0\textwidth]{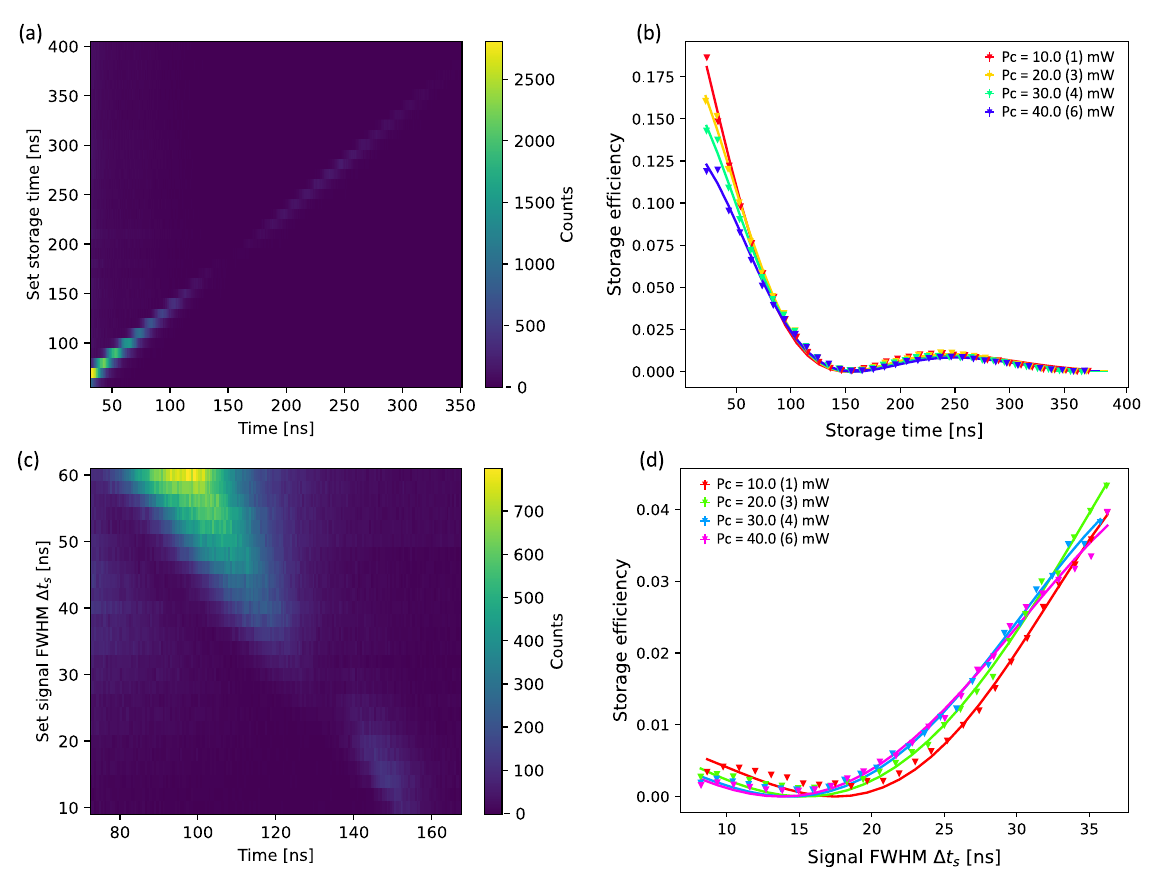}
    \caption{\textbf{Memory efficiency as a function of lifetime and bandwidth.} \textbf{(a)} Heat map of the intensity of the read signal for varying set storage times with $P_c = 10.0(1)~mW$, $\Delta t_{s} = 14.1(4)~ns$. An oscillatory behavior due to spin precession is observed. \textbf{(b)} Storage efficiency as a function of storage time for different control powers. The internal efficiency reduces with increasing powers for shorter storage times. \textbf{(c)} Heat map of the intensity of the read signal with varying input signal pulse widths, $\Delta t_s$, with a fixed set storage time of $150~ns$. \textbf{(d)} Storage efficiency as a function of the signal pulse width for different control powers. The storage bandwidth increases with increasing control power. 
    }
    \label{fig-lifetime_bandwidth}
\end{figure}

As a next step, we investigated the influence of the control power $P_c$ on memory efficiency. The heat map in Fig. \ref{fig-lifetime_bandwidth} (a) shows the intensity of the read signal (retrieved pulse after storage) as a function of the measured and actual set storage times for a signal pulse width of $\Delta t_{s} = 14.1(4)~ns$ and $P_c = 10.0(1)~mW$. Again, the oscillatory behavior due to spin precession is observed. From the diagonal cross sections of such heat maps for different control powers the storage efficiency $\eta_\mathrm{int}$ as a function of the storage time is derived and plotted in Fig. \ref{fig-lifetime_bandwidth} (a). Counterintuitively, $\eta_\mathrm{int}$ decreases with increasing $P_c$ for shorter storage times. This can be attributed to the power-dependent spatial compression of the input signal pulse while traveling through the LC in the atomic vapor cell. With increasing control power, the EIT transparency window broadens, but the group velocity decreases. Following references \cite{davidson-marquis_coherent_2021,fleischhauer_electromagnetically_2005}, we simulated the group velocity of a pulse propagating through the cell (see supplementary information \ref{SI-EIT-simulation}). At $Pc = 10~mW$, a pulse compression by a factor of 143 occurs, i.e., a 14 ns signal pulse traveling through the LC is compressed to a length of 29 mm. As the optical length of the LC is 5 mm, only a fraction of the pulse is stored. Increasing $P_c$ reduces the compression further and leads to even longer signal pulses in the LC. For this reason, increasing $P_c$ reduces the storage efficiency in our scenario. This effect becomes less pronounced for longer storage times where the intrinsic loss mechanisms, such as decoherence by atomic collisions, dominate. To obtain a higher efficiency the LC has to be extended or the optical depth has to be increased. 

Finally, we characterize the bandwidth of the LC-memory by setting a fixed storage time and varying the signal and control pulse widths, where we keep the ratio of the control to the signal width constant, i.e., $\Delta t_c=1.5\Delta t_s$. The results are shown in Fig. \ref{fig-lifetime_bandwidth} (c)-(d). The shift between measured time and set storage time arises from the changes in the control pulse width. Broader control pulses result in earlier read pulses (see supplementary information \ref{SI-time-ref}). An important parameter of a memory is its bandwidth. It is defined as $\Delta f_\mathrm{BW}=1/\Delta t_s$ for a value of $\Delta t_s$ where the storage efficiency $\eta_\mathrm{mem}$ is -3dB and can be obtained from the fits. Here, we find a bandwidth of $35.2(6)~MHz$. These results indicate that the optimal operating point for the LC-memory is found in the low $P_c$ regime. Importantly, we can utilize higher power levels than necessary without compromising the transmission through the LC. This capability paves the way for investigating non-linear processes that require high-intensity light fields guided within the LCs without causing damage.

\begin{figure}[ht]
    \centering
    \includegraphics[width=0.7\textwidth]{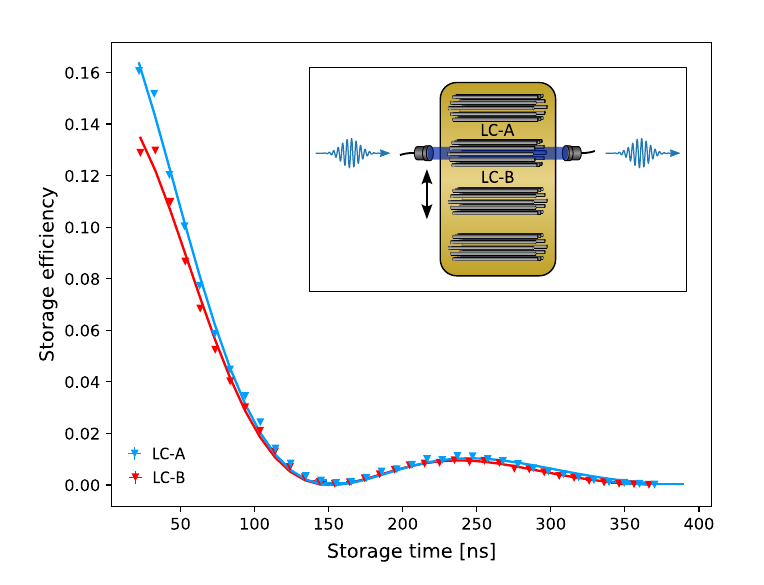}
    \caption{\textbf{Storage of light in two different LCs on a chip.} Storage efficiency measurements were performed in two neighboring LCs with the same geometrical parameters (LC-A and LC-B) on a single chip. Control powers and signal pulse widths were $P_c=20.0(3)~mW$ and $\Delta t_{s}=14.2(4)~s$, respectively. Both LCs perfom nearly identically within the fitting errors, $t_{mem-A}=86(3)~ns$ and $t_{mem-B}=87(3)~ns$. {\bf Inset:} Schematic of the coupling geometry.}
    \label{fig-LC-A_LC-B}
\end{figure}

\subsection{Spatial multiplexing of light cage memories on a chip}

As indicated in Fig. \ref{fig-concept} spatial multiplexing requires the fabrication and integration of several identical quantum memories. We demonstrate this conceptually by printing several LCs on the same chip, only hundreds of $\mu m$ apart, and storing light pulses therein. The nanoprinting process allows precise and reproducible fabrication of complex on-chip hollow-core waveguide structures. Extensive studies show extraordinary high intra-chip reproducibility with structure variations of less than 2 nm and minimal inter-chip variations of less than 15 nm \cite{burger_ultrahigh-aspect-ratio_2021}, both of which is crucial for the multiplexing concept discussed here. As an example, the storage efficiency at a fixed pulse width of $14.1~ns$ was measured as a function of the storage time for two different LCs, labeled LC-A and LC-B, see Fig. \ref{fig-LC-A_LC-B}. The performance of both LC-A and LC-B are identical with the same storage times $86(3)~ns$ and $87(3)~ns$, respectively. The small difference in the storage efficiency is likely due to a different lens coupling efficiency of the signal and control fields to the LC. This shows that the employed nanoprinting technique creates LCs with replicable optical properties, laying the groundwork for spatial multiplexing using an array of LC-memories on the same chip.

\section{Discussion}\label{disc}

Nanoprinted hollow-core light cages in an atomic vapor environment are suitable to store light using EIT for hundreds of nanoseconds. A figure of merit to compare our LC-memory to other hollow-core light guiding structures is the fractional delay $F$. It is defined as the ratio of storage time to signal pulse width, $F=t_\mathrm{storage}/\Delta t_s$, and can be used to evaluate the performance of the memory. In our system, we achieved a fractional delay close to 4 for pulses with 14 ns temporal width. 
This is comparable to results obtained with hollow-core photonic crystal fibers (HCPCFs). Peters et al. \cite{peters_single-photon-level_2020} achieved $F\approx3$ using Rubidium atoms from a magneto-optic trap, while Sprague et al. \cite{sprague_broadband_2014} and Rowland et al. \cite{rowland_high-bandwidth_2024} both reached $F\approx10$ with $\lambda$- and ladder-systems, respectively, where the latter is suitable for broader bandwidth. We like to point out that our systems utilize warm atomic vapor which fills the LCs immediately, while HCFs need years to reach filling saturation. Further, the design of hollow-core fibers is limited by the fiber-pulling process, whereas the printing of LCs by two-photon polymerization allows for arbitrary three-dimensional shapes and rapid integration on a chip. The alumina-coated structure is well suited to withstand the reactive atmosphere of Cs showing no degradation of their transmission even after 5 years of being sealed in their vapor cells. 

We observed that the LC-memory preserves the spin coherence of the atomic vapor in its interior long enough to observe spin precession in residual magnetic fields. With external magnetic field coils, this can be precisely controlled \cite{finkelstein_practical_2023} and utilized to fine-tune the polarization of the stored light pulse. Such feature helps to improve filtering, but also is crucial for constructively interfering stored and retrieved light pulses as recently demonstrated in efficiency enhancement via light-matter interference by Burdekin et al \cite{burdekin_enhancing_2024}.  

The reproducibility of the storage performance of different LCs with the same geometrical parameters together with the small footprint of warm atomic vapor systems opens the possibility of spatial multiplexing with several LC-quantum memories  on the same chip. Exploiting the field confinement inside the LC core, together with nanoprinted fiber-coupling technologies developed for these structures \cite{jang_fiber-integrated_2021}, as well as fiber feedthroughs into vapor cells \cite{weller_high_2017} could scale up quantum memory integration and technology to a new level. 

Despite the advantages of the presented platform, the LC-memory still needs improvement to surpass macroscopic vapor cells in terms of internal efficiencies and fractional delays. The measured optical depth $OD_\mathrm{EIT} \approx 1$ is below the expected OD compared to the free space scenario. By measuring the optical depth for a focused beam in a 5 mm vapor cell, without LCs, we obtain $OD_\mathrm{EIT}=3.5(5)$ indicating the atomic density inside the LCs is effectively reduced, most probably due to the adsorption of Cs atoms on the walls of the waveguide. This could be compensated by increasing the temperature, but the current structure has a measured critical temperature of $80~\degree C$. Beyond this point, the coupling efficiency of the structure degrades rapidly due to the inelastic deformation of the polymer at higher temperatures. Increasing the length of the LCs and employing light-induced atomic desorption techniques (LIAD) \cite{sprague_broadband_2014,kaczmarek_ultrahigh_2015} would increase the optical depth in the waveguides significantly without the need to increase the temperature. 

The transmission properties of the light cage are crucial for the multiplexing concept presented here. The optical losses for the fabricated structure are approximately 1 dB/mm \cite{burger_ultrahigh-aspect-ratio_2021}, which is acceptable for this study but may need to be reduced if longer structures are considered. Future optimization strategies will focus on minimizing waveguide roughness, which currently limits further loss reduction. These strategies include annealing and advanced writing techniques, such as grayscale lithography \cite{aderneuer_two-photon_2021}. Additionally, increasing the air-filling fraction through local structuring, as shown in \cite{kim_locally_2022-1}, could further mitigate the impact of polymeric surfaces and enhance performance.

Another key aspect of advancing this concept is the photonic integration of structures into on-chip and fiber architectures. Recent advances in the field include fiber-coupled integration of light cages on V-groove functionalized silicon chips for spectroscopy \cite{jang_fiber-integrated_2021} and direct printing of light cages onto fiber end faces \cite{huang_fiber-interfaced_2024-1}.

In addition to the light cage geometry, a recently developed anti-resonant hollow core waveguide features a closed square polymer membrane surrounding the central hollow core gaps every 150 µm for gas access. This structure can be integrated into on-chip environments \cite{burger_3d-nanoprinted_2022-1} or interfaced with optical fibers and offers broad spectral transmission bands with reduced structural complexity, potentially benefiting future multiplexing experiments. Moreover, these waveguides can be interfaced with phase plates that allow selective mode excitation within the waveguide core \cite{pereira_spatially_2024}, further enhancing the flexibility of the concept.

In conclusion, we have achieved storage of faint light pulses in a LC structure for hundreds of nanoseconds for the first time. The fractional delay, a figure of merit of a light memory, is on par with other memories using light confinement. We demonstrated the versatility and reproducibility of the fabrication process by integrating several LC-memories on a single chip inside a Cs vapor shell. We found nearly identical storage behavior of individual LC. Our results can lead to a plug-and-play on-chip quantum memory or quantum synchronizer \cite{davidson_single-photon_2023} with multiplexing capabilities.

\section{Methods}\label{methods}

\textbf{Light Cage: fabrication and integration into an alkali vapor cell.} The LC was fabricated by 3D two-photon polymerization lithography on a silicon substrate using a commercial 3D printing system (Photonic Professional GT, Nanoscribe GmbH) and a negative resist (IP-Dip, Nanoscribe GmbH). A 63× objective with NA = 1.4 allowed a sub-micron lateral resolution of ~400 nm. A custom General Writing Language (GWL) program was used to optimize implementation conditions to improve mechanical stability. The lateral position of the laser beam was controlled by galvo mirrors at a speed of 55 mm/s. The system operated with a 780 nm NIR femtosecond laser (100 fs pulse width, 80 MHz repetition rate) and an average power of 12.4 mW before entering the objective. The horizontal layer spacing was 150 nm, with 100 nm between lines within a layer.  
After printing, the structure was coated with alumina using low-temperature ALD at a growth rate of 2.2 Å per cycle at 30°C. The chip was mounted on aluminum and secured with a fused glass clamp. The custom optical-quality glass cell, filled with less than 1 g of cesium under high vacuum, was placed in an oven with resistance heaters and insulated with a polyoxymethylene (POM) housing.

\textbf{EIT and light storage setup.} EIT is created using co-propagating orthogonally polarized signal and control fields, as shown in Fig. \ref{fig-setup}. Both lasers are locked to their atomic transition through saturated absorption spectroscopy (SAS). The control laser (DL pro and BoosTA, Toptica) undergoes $-50 dB$ suppression through high-quality calcite prisms and additional $-60~dB$ via cascaded Fabry-Pérot etalons with an FWHM of $400~MHz$ and $800~MHz$. The fundamental mode $\mathrm{TEM_{00}}$ is effectively coupled into the LC using the lenses $f$ (A220TM-B, Thorlabs) and inspected with a CMOS camera (Fig. \ref{fig-concept} (c)). The LC coupling achieves transmission efficiency $\eta_{LC}=0.20(1)$, with modal attenuation $\alpha=-1.51(2)~dB/mm$ at 894 nm, including losses due to the modal attenuation, coupling efficiency, and losses on the vapor cell windows. EIT spectra characterization in the LC is performed by scanning the signal laser (EYP-DFB-0894, Eagleyard Photonics) and the fiber-coupled photodiode (DET36A2, Thorlabs) for detection. The storage sequence is executed using a pair of EOMs (AM905b, Jenoptik) with active BIAS stabilization through optical power monitoring. The signal is detected with SNSPDs (Eos CS, Single Quantum) and analyzed with a TCSPC system (PicoHarp 300, PicoQuant).

\backmatter

\bmhead{Supplementary information}

Supplementary information is added as appendices A to C in this manuscript.

\bmhead{Acknowledgements}

This work was supported by the German Research Foundation (DFG), projects SCHM2655/15-1, SCHM2655/22-1, SCHM2655/21-1, BE2224/19-1, and the Federal Ministry of Education and Research (BMBF), project 16KISQ003. The authors thank Frank Scheiber for fabricating the glass cell.

\newpage

\begin{appendices}

\section{Electromagnetically induced transparency in lossy waveguides}\label{SI-EIT-simulation}
The spectra measured of light coupled into a light cage show spectral features absent in a focused free space scenario (see main text Fig. \ref{fig-EITspectra}). These additional "shoulder-like" features can be explained considering an exponentially decaying Rabi frequency of the control field $\Omega_c(z)$, as treated in Ref. \cite{davidson-marquis_coherent_2021}. The maximum Rabi frequency $\Omega_0$ is achieved at the focal spot of the lens used to couple the light into the LC. This value decays exponentially as the light propagates through the LC with a modal attenuation $\alpha$, i.e., $\Omega(z) = \Omega_0\,10^{-\alpha z/20dB}$, with a measured $\alpha=1.51(2)~dB/mm$ in our experiment. 

The electrical susceptibility of the atomic media under EIT at a specific position $z$ is given by the expression \cite{fleischhauer_electromagnetically_2005}: 

\begin{equation}
\begin{split}  
    \chi(\Delta\nu_s,z)=\frac{|\mu_{31}|^2\, \varrho}{\epsilon_0\, \hbar}\left[ \frac{4\,\Delta\,(\Omega_c^2(z)-4\,\Delta^2-\gamma_d^2)}{|\,\Omega_c^2(z)+(\gamma_{31}+i\,2\,\Delta)(\gamma_{d}+i\,2\,\Delta)\,|^2}\right.\\
    +\left.i\,\frac{8\,\Delta^2\,\gamma_{31}+2\,\gamma_{d}\,(\Omega_c^2(z)+\gamma_{31}\,\gamma_d)}{|\,\Omega_c^2(z)+(\gamma_{31}+i\,2\,\Delta)(\gamma_{d}+i\,2\,\Delta)\,|^2} \right].
\end{split}
\label{eq:chi}
\end{equation}    

Then, we calculate the measured transmission spectra of the light coupled into the LC in the alkali atmosphere by integrating the susceptibility along the length $L$ of the vapor cell,

\begin{equation}
    T(\Delta\nu_s)=\mathrm{exp}\left[ -\frac{4\pi\nu_0}{c_0}\, \mathrm{Im}\left\{\int_{0}^{L} \tilde\chi(\Delta\nu_s,z)\,dz\right\}\right],
\label{eq:trans}
\end{equation}

where $\tilde\chi(\Delta\nu_s,z)$ is the Doppler broadened susceptibility of the atomic ensemble, calculated as the convolution of the susceptibility $\chi(\Delta\nu_s,z)$ with a Maxwell-Boltzmann distribution of the atomic velocities given a mean thermal velocity \cite{figueroa_decoherence_2006}. It is important to note that the atomic density $\varrho$ also depends on the temperature of the gas in the cell and is calculated following Ref. \cite{siddons_absolute_2008}. The maximum control Rabi frequency $\Omega_0$ is left as a fitting parameter, as well as the temperature, to account for the reduction of atomic density inside the LC.

\begin{figure}[ht]
    \centering
    \includegraphics[width=1.0\textwidth]{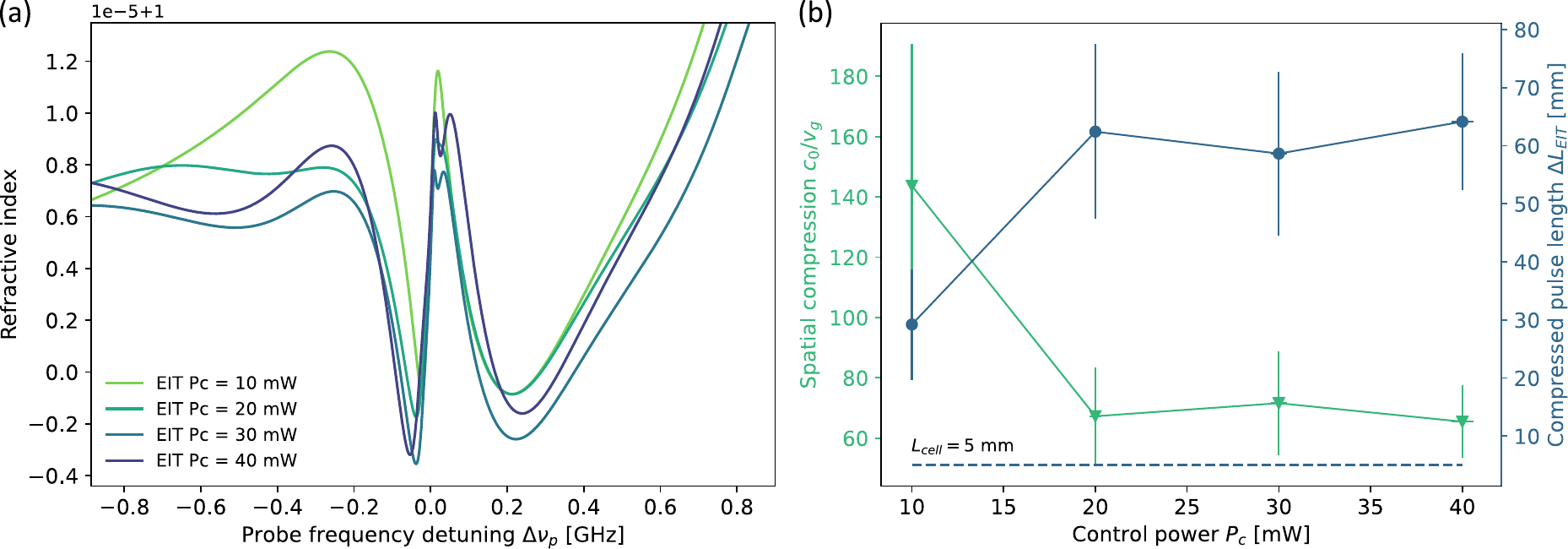}
    \caption{
    \textbf{Spatial pulse compression in the LC} (a) Simulated refractive index under the measured EIT conditions in the LC. For increasing control powers ($P_C$), the steepness of the slope on resonance is reduced. (b) Spatial compression factor $c_0 /v_g$ and the compressed length of a pulse after propagation through the LC as a function of $P_C$, for a signal pulse width $\Delta t_s=14~ns$. Only the portion of the signal pulse that fits the LC and vapor cell (dashed line) will be stored in the atomic media. 
    }
    \label{SI-fig-compression}
\end{figure}

From the susceptibility $\tilde\chi(\Delta\nu_s,z)$, we can compute the dispersion of a light pulse through the atomic media using via the refractive index, $n=\operatorname{Re}(1+\chi)$, in particular, the modification of the group velocity under EIT at the two-photon resonance is \cite{fleischhauer_electromagnetically_2005}:

\begin{equation}
    v_g=\left.\frac{c}{n+\nu_p\frac{dn}{d\nu_p}}\right|_{\Delta\nu_p=0},
\label{eq:vgroup}
\end{equation}

with $c$ the speed of light in vacuum. From this equation, the spatial compression of a propagating light pulse can be calculated as $L_{\mathrm{EIT}} = L_0\,v_g/c$. Utilizing the linear susceptibilities $\tilde\chi(\Delta\nu_s,z)$ obtained from the fitted transmission spectra, main text Fig. \ref{fig-EITspectra}, we compute the refractive index $n$, shown in Fig. \ref{SI-fig-compression}-(a). At the two-photon resonance ($\Delta\nu_p=0$), a positive slope that decreases with increasing control power $P_C$ is observed as expected. The spatial compression of the signal light pulse as a function of $P_C$ is computed using Eq. \eqref{eq:vgroup} and displayed in Fig. \ref{SI-fig-compression}-(b). Notably, the compressed pulses have a spatial extension considerably larger than the vapor cell, explaining the limited efficiency considering the achieved optical depth, both crucial parameters for optimal storage \cite{gorshkov_optimal_2008}. Furthermore, increasing $P_C$ increases the memory bandwidth at the cost of spatial compression, reducing the memory efficiency for higher bandwidths, as observed in Fig. \ref{fig-lifetime_bandwidth}-(b), -(d). An increase in the optical depth is required to achieve greater efficiency at high bandwidths.

\section{Magnetic field influence on the storage efficiency}\label{SI-Bfield-efficiency}

The decrease of the efficiency of a quantum memory as a function of the storage time can be modeled using an exponential with decay constant $t_\mathrm{mem}$ \cite{phillips_storage_2001,katz_light_2018}, the so-called memory lifetime. In Fig. \ref{fig-storage_analysis} and Fig. \ref{fig-lifetime_bandwidth} of the main text, we can see an apparent deviation from the expected exponential decay in the form of a damped oscillation of the storage efficiency storage. 

The reason is that light is stored as a spin-wave excitation in the atomic vapor, which makes the memory sensitive to spin precession caused by stray magnetic fields. These oscillations result in a change to the initially linearly polarized signal light after storage. However, since we employ polarization filtering, a portion of the signal is filtered out and goes undetected. Residual magnetic fields are expected since the vapor cell has no magnetic shielding. To extract $t_\mathrm{mem}$ from the measured data, a heuristic fit that accounts for the precession is used in the form of an exponentially damped oscillation of the efficiency with a corresponding Larmor frequency of the electronic spins, $\omega=\gamma B$ (with $\gamma$ the electron gyromagnetic ratio): 

\begin{equation}
    \eta_\mathrm{int}=e^{-t/t_\mathrm{mem}} \mathrm{sin}^2 \left( \frac{\omega}{2} t+\phi \right)
    \label{eq-lifetime}
\end{equation}

The fit and its 95\% confidence interval are shown in Fig. \ref{fig-storage_analysis}-(b) in the main text, by having $t_\mathrm{mem}$, $\omega$ and an arbitrary phase $\phi$ as fitting parameters. There is excellent agreement with the experimental data, and we find a memory lifetime of $t_\mathrm{mem} = 84(2)~ns$. The spin precession frequency obtained from this model is $f=3.573(2)~MHz$, corresponding to a magnetic field of $127.49(5)~\mu T$. This agrees with the experimental conditions, considering the Earth's magnetic field intensity around $50~\mu T$ \cite{alken_international_2021}. In addition, the vapor cell is heated using a pair of resistors with a current of $4~A$, inducing an estimated magnetic field of $80~\mu T$ in the region of the LC chip. The observed spin precession offers an advantageous way to finely control the polarization of the stored light pulse. Magnetic polarization control of stored light in a quantum memory will thus be a subject of further study.

\section{Storage sequence of light pulses}\label{SI-time-ref}

\begin{figure}[ht]
    \centering
    \includegraphics[width=1.0\textwidth]{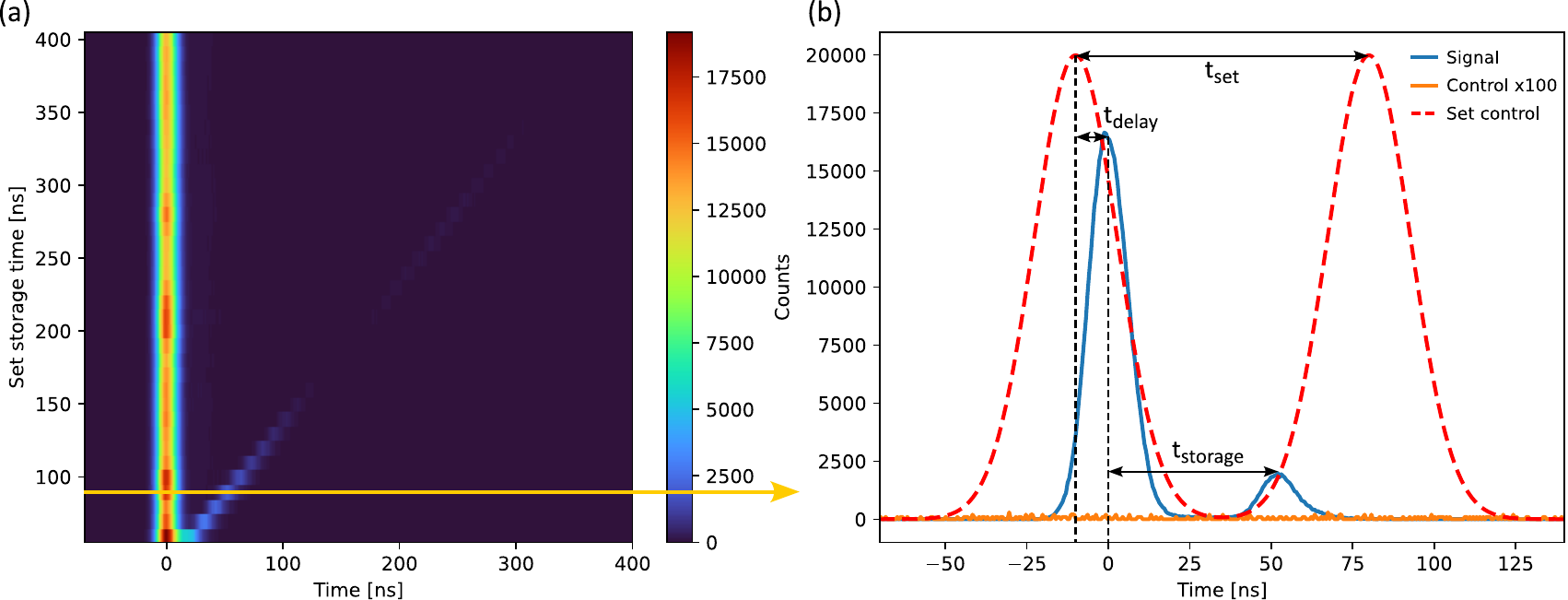}
    \caption{
    \textbf{Set and measured storage times.} \textbf{(a)} Heat map of the intensity of the detected signal for varying storage times with $P_c = 10.0(1)~mW$, $\Delta t_{s} = 14.1(4)~ns$. \textbf{(b)} Example of an individual measurement for a set storage time $t_\mathrm{set}=90~\mathrm{ns}$. The signal without noise correction comprises the leak and read pulses (blue curve), where the center of the leak is used as the origin of the time scale. The control field intensity (orange curve), here scaled by a factor of 100, evidence the high signal-to-noise ratio achieved. A reconstruction of the control pulses (dashed curve). $t_\mathrm{delay}$ corresponds to the set delay between the write pulse and input signal, $t_\mathrm{set}$ is the set time between write and read pulses, $t_\mathrm{storage}$ the measured storage time.
    }
    \label{SI-fig-timings}
\end{figure}

The detected intensities and arrival times for varying set storage times are shown in Fig. \ref{SI-fig-timings} (a). Here, an offset between the measured storage time ($t_\mathrm{storage}$) and the set storage time ($t_\mathrm{set}$) is visible. This effect arises from the storage sequence utilized in our experiments. The control field consists of a pump pulse followed by two Gaussian pulses of full width at half maximum (FWHM) $\Delta t_c$, labeled write and read pulses, as shown in Fig. \ref{SI-fig-timings}-(b). The input signal is also a Gaussian pulse with FWHM $\Delta t_s$, following a fixed ratio to the width of the control pulses, $\Delta t_s=(2/3)\,\Delta t_c$. The signal is synchronized to arrive with a delay $t_\mathrm{delay}$ from the center of the write pulse. The highest read intensity is achieved by letting the signal arrive on the latter part of the write pulse. This delay was found to be $t_\mathrm{delay}=(1/3)\,\Delta t_c$. Due to the time-reversal condition for optimal retrieval \cite{gorshkov_universal_2007}, this results in a retrieved pulse arising on the leading part of the read pulse. The actual storage time, or measured storage time ($t_\mathrm{storage}$), is therefore shorter than the set storage time, i.e., $t_\mathrm{storage}<t_\mathrm{set}-(2/3)\,\Delta t_c$.

In Fig. \ref{SI-fig-timings}-(a), the control width is kept constant while the set storage time gradually increases, resulting in a continuous shift between $t_\mathrm{set}$ and $t_\mathrm{storage}$. As an example, we take the measured histogram at $t_\mathrm{set} = 90~\mathrm{ns}$ with a measured storage time of 52.37(4) ns, see Fig. \ref{SI-fig-timings}-(b). In the case of constant $t_\mathrm{set}$ with increasing $\Delta t_s$, this discrepancy with the measured storage time becomes more evident as wider control pulses require longer $t_\mathrm{delay}$, resulting in progressively shorter storage times, see main text Fig. \ref{fig-lifetime_bandwidth}-(c).

\end{appendices}

\newpage

\bibliography{sn-article}

\end{document}